\documentclass{PoS}

\newcommand{\be}{\begin{eqnarray}}
\newcommand{\ee}{\end{eqnarray}}

\def \a               {\alpha}
\def \ti              {\tilde}
\def \st              {\ti t}
\def \sb              {\ti b}
\def \sg              {\ti g}

\title{CP violation in associated production of a charged Higgs boson and a top quark at the LHC}

\ShortTitle{CP violation in $H^\pm t$ at the LHC}

\author{\speaker{Elena Ginina}\\
        Institut f\"ur Hochenergiephysik der \"Osterreichischen Akademie der
     Wissenschaften,  A-1050 Vienna, Austria\\
        E-mail: \email{eginina@hephy.oeaw.ac.at}}

\author{Ekaterina Christova$^{1}$, Helmut Eberl$^{2}$\\
        \small $^{1}$Institute for Nuclear Research and Nuclear Energy, BAS, Sofia 1784, Bulgaria\\
    $^{2}$Institut f\"ur Hochenergiephysik der \"Osterreichischen Akademie der
     Wissenschaften,  A-1050 Vienna, Austria\\
        E-mails: \email{echristo@inrne.bas.bg, helmut@hephy.oeaw.ac.at}}

\abstract{We study effects of CP violation in an associated
production of a charged Higgs boson and a top quark at the LHC:
$pp \to tH^\pm +X$. We calculate the CP violating asymmetry
between the total cross section for $H^+$ and $H^-$ production at
next-to-leading order in the MSSM, and perform a detailed
numerical analysis.  In the production only the asymmetry is of
the order of 20\%. The asymmetry in the production and any
subsequent decay of an on-shell charged Higgs boson is to a good
approximation the sum of the asymmetry in the production and the
asymmetry in the decay. We consider subsequent decays of $H^\pm $
to $tb$, $\nu_\tau \tau^\pm$ and $Wh^0$. In the case of subsequent
$H^\pm \to tb$ decay, the $W^\pm - H^\pm$ self energy
contributions from the production and the decay cancel. However,
the remaining effect, mainly due to CP violating box graphs with
gluino can go up to $\sim 13$\%.}

\FullConference{Prospects for Charged Higgs Discovery at Colliders\\
             September 16-19 2008\\
             Uppsala, Sweden}

\begin{document}

\section{Preface}

Recently we studied effects of CP violation (CPV) in the decays of the
charged Higgs boson into ordinary particles: $H^\pm \to t b$, $H^\pm \to
\nu\tau^\pm$ and $H^\pm \to W^\pm h^0$ in the MSSM~\cite{Hplustb, Hplustaunu, HplusWh, Hdecays}. Loop corrections induced by a
Lagrangian with complex coupling parameters lead to non zero decay rate asymmetries between the partial
decay widths of $H^+$ and $H^-$. We found that in the $H^\pm \to tb$ decay such effects can be rather large and the asymmetry
can go up to $\sim 25$\%~\cite{Hplustb}. This motivated our interest in studying CPV in the production of $H^\pm$ at
the LHC~\cite{Jennifer,jung:lee:song}, where the dominant  production process is associated with a top quark production:
$pp\to H^\pm t +X$. As we are interested in mass range $m_{H^+}\ge 400$ GeV, at parton level we only consider bottom-gluon
fusion: $bg\to H^\pm t $. The latter process contains the same $H^\pm tb$ vertex and corresponding loop diagrams as the
decay $H^\pm \to tb$, and one would expect that the CPV effects  might be of the same magnitude. Moreover, in the
production process there are box graphs, that are of the same order and must be taken into account as well, as additional
source of CPV.

After the charged Higgs is produced in proton-proton collisions, it
will be identified through some of its decay modes. We
study CPV in the combined process of $H^\pm$ production and decay,
considering $H^\pm$ decays into $t b$, $\nu\tau^\pm$ and $ W^\pm h^0$.

\section{The subprocess $bg \rightarrow t H^\pm$}

We have the following processes, connected by charge conjugation:
\begin{eqnarray}
b_r (p_b) + g_\mu^\a (p_g) \longrightarrow  t_s (p_t) + H^-
(p_{H^-}) \label{pro}\,,~\\ {\bar b}_r (p_{\bar b}) + g_\mu^\a
(p_g) \longrightarrow  {\bar t}_s (p_{\bar t}) + H^+ (p_{H^+})
\label{procon}\,,~
\end{eqnarray}
where $r, s $ and $ \a$ are colour indices, $r,s=1,2,3; \a=1,...,8.$
\begin{figure}[h!]
 \begin{center}
 \mbox{\resizebox{!}{2.5cm}{\includegraphics{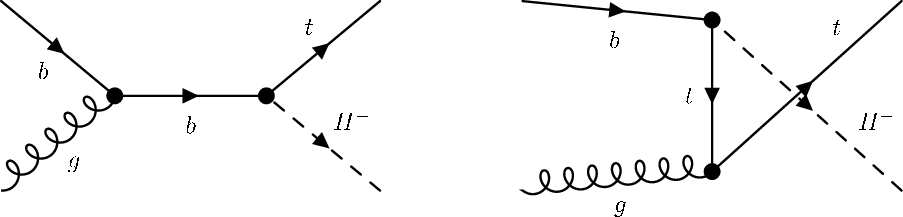}}}
 \vspace*{-8mm}
  \end{center}
  \caption {The tree level graphs of the $b g \to t H^-$ process.}
  \label{tree}
\end{figure}

The tree-level process (\ref{pro}) contains two graphs (Fig.~\ref{tree}): with bottom-quark exchange ($s$-channel),
and top-quark exchange ($t$-channel).
The Mandelstam variables are: $\hat s=(p_b+p_g)^2,~ \hat t=(p_t-p_g)^2=(p_b-p_{H^-})^2.$

\section{The LHC process: $pp \rightarrow t H^\pm +X$}

We consider charged Higgs boson production in proton-proton collisions:
\begin{equation}
p\, (P_A)+p\, (P_B)\rightarrow t(p_t)+H^\pm(p_{H^\pm})+X. \label{hadronprocesses}
\end{equation}
The Mandelstam variable is: $S =(P_A+P_B)^2$ ( for LHC $\sqrt{S}=14 $
TeV) and we set: $p_b= x_b P_A=\tilde{x}_b P_B$ and $p_g=x_g
P_B=\tilde{x}_g P_A$, where $x_i$ ($\tilde{x_i}$)\,is the momentum
fraction of the hadron $B (A)$ carried by the parton $i$.
Neglecting the proton mass compared to $\sqrt{S} $, we get
$\hat{s}= x_b x_g S=\tilde{x}_b \tilde{x}_g S$.

We define the CPV asymmetry in the $H^\pm$ production as the difference between the total number of produced
$H^+$ and $H^-$ in proton-proton collisions:
\begin{equation}
A_P^{CP}={\sigma(pp\rightarrow \bar{t}H^+) -\sigma(pp\rightarrow t
H^-)\over \sigma(pp\rightarrow \bar{t}H^+) +\sigma(pp\rightarrow t
H^-)}\,, \label{hadronasymm}
\end{equation}
where the total cross sections in (\ref{hadronasymm}) are given by:
\begin{equation}
\sigma^\pm=\sigma (pp\rightarrow \bar t H^+, t H^-)=2\int_0^1 f_b(x_b)\int_0^1
f_g(x_g)\hat{\sigma}^\pm (x_b x_g S)\theta (x_b x_g S-S_0)dx_b
dx_g.
\label{account}
\end{equation}
Here $\hat \sigma^\pm$ are the parton level cross sections for $H^\pm$ production in bottom-gluon fusion,
$S_0=(m_t+m_{H^+})^2$ fixes the kinematically allowed energy range, $f_b$ and $f_g$ are the parton distribution functions
(PDF's) of the bottom and the gluon in the proton,
$f_{ b}(x_{ b})= f_{\bar b}(x_{\bar b})$, and the factor 2 counts the two possibilities: $b\ (g)$ comes from the proton
$A\ (B)$ and
${\it vice}$ ${\it versa}$.

The CPV asymmetry (\ref{hadronasymm}) is caused by loop corrections with complex coupling parameters. There are three types of
MSSM loop corrections to both $s$- and $t$-channels that contribute in $A^{CP}_P$: corrections to the
$H^\pm tb$-vertex, selfenergy loops on the $H\pm$-line and box diagrams~\cite{progresswork}. The total one-loop cross
sections of the processes (\ref{hadronprocesses}) have CP invariant and CP violating parts:
$\sigma^\pm=\sigma^{inv} \pm \sigma^{CP}$, and for the asymmetry, up to terms linear
in $\alpha_s$ and $\alpha_w$, we obtain~\cite{progresswork}:
\begin{eqnarray}
 A_P^{CP}=\frac{\sigma^{CP}}{\sigma^{tree}}\,.
\end{eqnarray}

\section{$H^\pm$ production and decay at the LHC}

We define the CPV asymmetry in charged Higgs boson production in $pp \to tH^\pm$, with a subsequent decay
 $H^\pm \to f$, asumming CPV in both production and decay, as:
\begin{equation}
A^{CP}_{f}={\sigma(pp\rightarrow \bar{t}H^+\to  \bar{t}f)
-\sigma(pp\rightarrow t H^-\to t \bar f)\over
\sigma(pp\rightarrow \bar{t}H^+\to  \bar{t} f)
+\sigma(pp\rightarrow t H^-\to t \bar f)}\,,
\label{panddasymm}
\end{equation}
where $f$ stands for the chosen decay mode:
$f=t\bar b;\, \nu\tau^+$ and $ W^+ h^0$.

In narrow width approximation, the asymmetry (\ref{panddasymm}) is an algebraic sum of the
CPV asymmetry $A^{CP}_P$ in the production,  and the
CPV asymmetry $A^{CP}_{D,f}$ in the decay $f$ of the charged Higgs boson\footnote{In~\cite{Hplustb, Hplustaunu, HplusWh, Hdecays}
and~\cite{Katqproceedings} the asymmetry $A^{CP}_{D}$ is denoted with $\delta^{CP}$.}:
\begin{equation}
A^{CP}_f= A_{P}^{CP}+A^{CP}_{D,f}. \label{finalf}
\end{equation}

\section{Numerical analysis}

We present numerical results for the charged Higgs rate
asymmetries $A^{CP}_P, A^{CP}_{t b}$, and $A^{CP}_{\nu \tau}$ in
the MSSM.
All formulas used in the numerical code are calculated analytically and can be found in~\cite{progresswork}, except for the box
contributions, which are rather lengthy. Furthermore, all
individual one-loop contributions are checked numerically using the
packages FeynArts and FormCalc~\cite{FeynArts}. We also use
LoopTools, see again~\cite{FeynArts}, and FF~\cite{FFpackage}.
In the numerical code the Yukawa couplings of the third generation
quarks ($h_t$, $h_b$) are taken to be running~\cite{Hplustb}, at the scale $Q =
m_{H^+} + m_t$. For the evaluation of $f_b$ and $f_g$ we
use CTEQ6L~\cite{PDFs}, with LO PDF's and NLO $\alpha_s$, at the
same scale $Q$. We assume GUT relation between $M_1$ and $M_2$, so
that the phase of $M_1=0$. Our numerical study showed that the
contribution of the loop diagrams with chargino, neutralino, stau
and sneutrino in the considered CPV asymmetries are
negligible and we show contributions from diagrams with $\st \sb$
and $ \sg$ only. If not specified otherwise, we fix the following
MSSM parameters: $ M_2=300~{\rm GeV},~M_3=727~{\rm GeV},~M_{\tilde
U}=M_{\tilde Q}=M_{\tilde D}=350~ {\rm GeV},~\mu=-700~{\rm
GeV},~|A_t| = |A_b| =700~{\rm GeV}, \tan \beta=5,~\phi_{A_t}=\pi/2,~\phi_{A_b}=\phi_{\mu}=0$. The relevant masses of the
sparticles for this choice of parameters, $\tan \beta = 5$ or
30 are shown in Table 1 of~\cite{Katqproceedings}. Our numerical results are in agreements with those in~\cite{Jennifer}, but
we disagree analytically and numerically with the results given in~\cite{jung:lee:song}.

\subsection{Production only}

\begin{figure}[h!]
\begin{center}
\mbox{\resizebox{76mm}{!}{\includegraphics{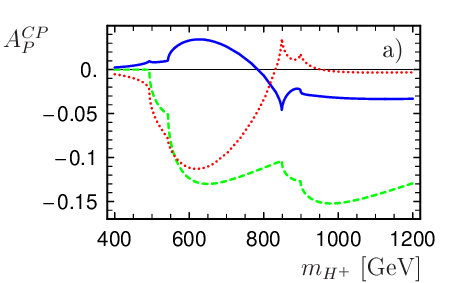}}
\resizebox{76mm}{!}{\includegraphics{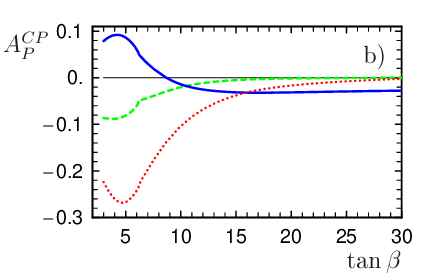}}}
 \vspace*{-13mm}
\end{center}
\caption{The various contributions to the asymmetry $A_P^{CP}$
at hadron level for the chosen set of parameters: ~a)~as a function of $m_{H^+}$; ~b)~ as a function of $\tan\beta$, $m_{\tilde g} =$~450~GeV,
$m_{H^+} =$~550~GeV. The red
dotted line corresponds to box graphs with gluino, the solid blue
one to the vertex graph with gluino, and the green dashed one to
the $W^\pm - H^\pm$ selfenergy graph with $\tilde t \tilde b$ loop.}
\label{fig1}
\end{figure}

As expected~\cite{Hplustb}, the CPV asymmetry in the production due to loop corrections
with $\st \sb $ and $\sg $ is of the same order of magnitude as in the case of the decay $H^\pm \to tb$, and can go up to
$\sim 20$\%. Moreover, the contributions of the box graphs that do not have an analog in the decay is significant and can
be dominant for relatively small $m_{H^+}$. The contributions of the vertex, selfenergy and box graphs with $\st \sb $ and $\sg$ to
the asymmetry $A_P^{CP}$ at hadron level as functions of $m_{H^+}$ are shown on Fig.~\ref{fig1}a. The large effect seen on
the figure is mainly due to the phase of $A_t$ and the asymmetry reaches its maximum for a maximal phase $\phi_{A_t}=\pi /2$.
The phase of $A_b$ doesn't have a big influence on the asymmetry and therefore we usually set it to zero.

The asymmetry $A_P^{CP}$ reaches its maximum value at $\tan \beta =5$ and falls down quickly with increasing $\tan \beta$.
This dependence for $m_{H^+}=550$ GeV is shown on Fig.~\ref{fig1}b.

\subsection{Production and subsequent decay}

\begin{figure}[h!]
\begin{center}
\mbox{\resizebox{76mm}{!}{\includegraphics{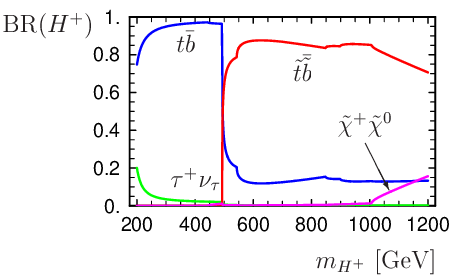}}}
 \vspace*{-8mm}
\end{center}
\caption {The tree-level branching ratios of $H^+$ for the chosen set of
parameters, as a function of $m_{H^+}$.}
\label{figBR}
\end{figure}

First we want to add a few remarks on the branching ratios (BR) of the relevant decays.
For small $m_{H^+}$, below the $\st \sb$ threshold, the dominant decay mode
is $H^\pm \to t b$, with BR $ \approx 1$,
while the BR of $H^\pm \to \nu \tau^\pm$ is in the order of a few percent, decreasing with increasing $m_{H^+}$. When the
$H^\pm \to \tilde t \tilde b$ channels are kinematically allowed, they start to dominate, and the BR of
$H^\pm \to \nu \tau^\pm$ to a good approximation becomes zero. However, the BR of  $H^\pm \to t b$ remains stable of the order
of 15-20\%, see Fig.~\ref{figBR}. The BR of $H^\pm \to W^\pm h^0$ reaches a few percent for small $\tan\beta$
in a relatively narrow range of
$m_{H^+}$~\cite{HplusWh}. In the considered range of parameters this decay is very much suppressed and we do not investigate it numerically.

\begin{figure}[h!]
\begin{center}
{\unitlength=1mm
\begin{picture}(140,48)(0,0)
\put(-4,0){\resizebox{74mm}{!}{\includegraphics{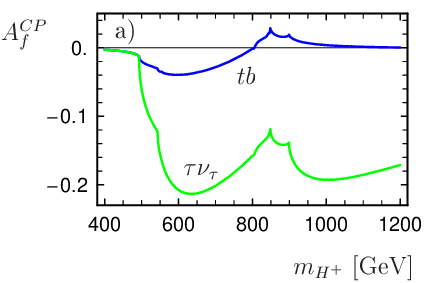}}}
\put(70,1.9){\resizebox{79mm}{!}{\includegraphics{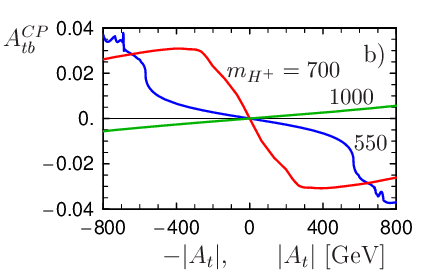}}}
\end{picture}
}
\end{center}
\vspace*{-6mm}
\caption {The total asymmetry $A^{CP}$ at hadron
level for the chosen set of parameters: ~a)~ as a function of
$m_{H^+}$. The blue line corresponds to the case when $H^\pm$
decays to $t b$, and the green one to $H^\pm$ decay to $\nu_\tau
\tau^\pm $; ~b)~ as a function of $|A_t|$, for three values of
$m_{H^+}$ (in GeV).} \label{fig3}
\end{figure}

In Fig.~\ref{fig3}a we show the total production and decay asymmetry $A^{CP}_f$ at hadron level, for $f= t b$ and $f= \nu\tau^\pm$.
Though for $H^\pm \to \nu \tau^\pm$ it can go up to $\sim 20$\% for $m_{H^+}\approx 650$ GeV, the BR of this decay in this
range of $H^+$ masses is too small and observation at LHC is impossible.

On the other hand, as the CPV asymmetries in the production and
the decay are additive, one can suppose that the total asymmetry
will be large. Moreover, in the case of $H^\pm \to tb$ decay alone
it is large~\cite{Hplustb, Katqproceedings}, with leading
contribution coming from the $H^\pm - W^\pm$ selfenergy graph with
$\tilde t \tilde b$ loop. In \cite{progresswork} we show
analytically that the $H^\pm - W^\pm$ selfenergy contribution to
the asymmetry $A^{CP}_{t b}$ of the decay part cancels exactly the
$W^\pm - H^\pm$ selfenergy contribution of the production part.
Our numerical study showed that the contributions of the vertex
graphs from the production and from the decay also partially
cancel with the box diagrams contribution. However, as the box
graphs do not have a real analogue in the decay, their
contribution remains leading in our studied case.

On  Fig.~\ref{fig3}b the dependence of $A^{CP}_{tb}$ on the absolute value of $A_t$ is shown for three different $m_{H^+}$.

\section{Summary}

We have calculated the CPV asymmetries $A^{CP}_P$, and
$A^{CP}_{f}$, with $f=tb;\, \nu\tau^\pm$ and $ W^\pm h^0$, between
the total cross sections of $H^+$ and $H^-$ production in
proton-proton collisions, proceeding at parton level through $b g$
fusion. We have performed a detailed numerical analysis, varying
the different parameters and phases of the MSSM. The asymmetry
$A_P^{CP}$ can go up to $\sim 20$\%  at $m_{H^+} \approx  600
~{\rm GeV}$, $\tan \beta=5$ and a maximal phase of $A_t$. This
effect is due to CPV vertex, selfenergy and box contributions with
$\st$, $\sb$ and $\sg$. The total asymmetry in the combined
process of production and a subsequent decay is approximately the
sum of $A^{CP}_P$ and $A^{CP}_{D, f}$, where $f$ is the relevant
decay. Despite the dominant CPV contribution from the decay
cancels with the relevant part of the production,  most promising
remains the $tb$ channel. The effect in this case is mainly due to
box diagrams with gluino and the asymmetry $A^{CP}_{tb}$ can go up
$\sim 13$\%~\cite{progresswork}.

\section*{Acknowledgments}

The authors acknowledge support from EU under the MR TN-CT-2006-035505 network
programme. This work is supported by the "Fonds zur F\"orderung der
wissenschaftlichen Forschung" of Austria, project No. P18959-N16.

\end{document}